\documentclass[10pt, a4paper]{article}
\usepackage{lrec2022} %
\usepackage{multibib}
\newcites{languageresource}{Language Resources}
\usepackage{graphicx}
\usepackage{tabularx}
\usepackage{soul}
\usepackage{titlesec}
\titleformat{\section}{\normalfont\large\bfseries\center}{\thesection.}{1em}{}
\titleformat{\subsection}{\normalfont\SmallTitleFont\bfseries\raggedright}{\thesubsection.}{1em}{}
\titleformat{\subsubsection}{\normalfont\normalsize\bfseries\raggedright}{\thesubsubsection.}{1em}{}
\renewcommand\thesection{\arabic{section}}
\renewcommand\thesubsection{\thesection.\arabic{subsection}}
\renewcommand\thesubsubsection{\thesubsection.\arabic{subsubsection}}

\usepackage{epstopdf}
\usepackage[utf8]{inputenc}

\usepackage{hyperref}
\usepackage{xstring}

\usepackage{color}
\usepackage{amsmath}

\usepackage{comment}

\newcommand{\us}{\rule{.4em}{.4pt}}

\title{Topic Segmentation of Research Article Collections}

\name{Erion Çano, Benjamin Roth} 

\address{Digital Philology \\
         Research Group Data Mining and Machine Learning \\
		 University of Vienna, Austria \\
         \{erion.cano, benjamin.roth\}@univie.ac.at\\}

\abstract{
Collections of research article data harvested from the web have become common recently since they are important resources for experimenting on tasks such as named entity recognition, text summarization, or keyword generation. In fact, certain types of experiments require collections that are both large and topically structured, with records assigned to separate research disciplines. Unfortunately, the current collections of publicly available research articles are either small or heterogeneous and unstructured. In this work, we perform topic segmentation of a paper data collection that we crawled and produce a multitopic dataset of roughly seven million paper data records. We construct a taxonomy of topics extracted from the data records and then annotate each document with its corresponding topic from that taxonomy. As a result, it is possible to use this newly proposed dataset in two modalities: as a heterogeneous collection of documents from various disciplines or as a set of homogeneous collections, each from a single research topic.     
\\ \newline  }

\begin{document}

\maketitleabstract

\section{Introduction}
\label{sec:intro}

The trend towards the online publication of research has created opportunities for crawling the web and producing research paper data collections. These datasets are being used to conduct experiments on text-related tasks such as text summarization, topic recognition or keyword generation \cite{nallapati-etal-2016-abstractive,8981519}. They were initially small, limited to a few thousand samples \cite{hulth-2003-improved,kim-etal-2010-semeval}. The recent ones have grown to millions of records in size and can be used to train larger data-driven predictive models such as deep neural networks \cite{cano-bojar-2019-efficiency,DBLP:journals/corr/abs-1804-08875,10.1145/3443279.3443305}.   
One problem with the existing paper data collections is that they are heterogeneous, mixing inside works from various scientific disciplines. This makes them useless when one wants to explore the role of topicality in text-related tasks (e.g., comparing the performance of data that are topically homogenous vs. heterogeneous).
One way to solve this problem is by performing topic segmentation on existing heterogeneous and topically unstructured collections. This can be achieved using topic modeling methods for finding a taxonomy of topics and then annotating each data sample with its most probable topic from the taxonomy. Following this approach, we create in this work OAGT, a paper data collection of 6\,942\,930 records intended for research.\footnote{\url{https://zenodo.org/record/6560535}} We combine two popular topic modeling methods, LDA (Latent Dirilecht Allocation) and Top2vec to analyze the data and construct a taxonomy of topics that are represented by their most significant terms.     
This way, we attach a topic to each of the data records, making it easy to recognize and utilize the resulting topically structured subsets separately for multitopic keyword generation or other types of experiments. 
In the near future, we will try to improve the topic representation of the dataset records by adding explicit annotations. To this end, we will try to integrate external knowledge from online publication libraries. Another possibility could be the manual annotation of a certain number of samples and the utilization of semi-supervised techniques for identifying the topic of all data collection samples \cite{10.1145/2983323.2983752,YAKIMOVICH2021100383}.  

\section{Background}
\label{sec:bckg}

\subsection{Scientific Data Sources} %
\label{ssec:sources}

Crawling the web for scientific publications and similar data is becoming a common practice that aims to provide valuable sources for research \cite{10.1145/1401890.1402008,info:doi/10.2196/17853}. Several authors are trying to exploit the web resources for building datasets that can be used for experiments in different text-related tasks.
\newcite{P17-1054} harvested computer science research articles and build KP20k, a popular dataset of 20 thousand data samples used for keyword generation. 
\newcite{DBLP:journals/corr/abs-1804-08875} collected papers from medicine (more specifically biomedical research) and released a large dataset of about 5 million articles. That collection is suitable for text summarization research (predicting paper abstract from the body, or title from the abstract).
\newcite{cano-bojar-2020-two} recently released two huge multitopic datasets, OAGSX of 34 million paper abstracts and titles and OAGKX of 23 million abstracts, titles, and keywords. The former is best suited for text summarization whereas the latter is more appropriate for keyword generation. The problem with both of them is the fact that the article topics are mixed, which makes it hard or impossible to run multitopic keyword generation or text summarization experiments.
This work aims to create a multitopic dataset with article data organized by topic. We build upon the Arnetminer initiative \cite{10.1145/1401890.1402008} and their OAG (Open Academic Graph) collection of academic works \cite{10.1145/3292500.3330785}. More specifically, we utilize OAG version 2.1\footnote{\url{https://www.aminer.cn/oag-2-1}} which is the latest, released in November 2020. The data processing and topic modeling steps are described in Section~\ref{sec:topicmod}.       

\subsection{Topic Modeling Research}
\label{ssec:topicmodel}

Research on automatic topic modeling and indexation has an early history that dates back to the nineties. \newcite{doi.org/10.1002} tried to solve the retrieval problem of matching query and document terms. They proposed LSI (Latent Semantic Indexing), a method that is based on matrix factorization (more specifically singular value decomposition) tries to uncover the latent semantic structure of documents. LSI was an important milestone that worked well in different retrieval and indexing tasks.        
\newcite{10.1145/312624.312649} proposed PLSI (Probabilistic Latent Semantic Indexing) which represents a significant advantage over LSI; solid foundations in statistical latent class model and maximization of log-likelihood function which minimizes word perplexity. Their experiments revealed gains of PLSI over LSI and standard term matching. Furthermore, PLSI can benefit from statistical methods of model fitting and combination.
LDA (Latent Dirichlet Allocation) is another popular probabilistic topic model that has been used for many years \cite{10.5555/944919.944937}. It is based on the exchangeability assumption (invariance of document order in collection and term order in a document), modeling documents as a finite mixture of topics, which on the other hand, are modeled using topic probabilities. Significant improvements are reported by comparing against PLSI. 
Despite the success of these above methods for many years, they do also exhibit limitations and drawbacks. In their foundation, they rely on the exchangeability assumption (bag-of-words representation) and ignore word order in documents. Moreover, to provide optimal results, they require certain preprocessing steps like removing stop-words, knowing the number of topics ahead of the modeling process, etc.  
The recent developments involving distributed representations of words \cite{10.5555/2999792.2999959} or longer text units like sentences and documents \cite{10.5555/3044805.3045025}, and especially pretrained language models \cite{devlin-etal-2019-bert}, have been proven very successful in capturing word semantics and context. They have opened opportunities to reframe topic modeling from different perspectives.
Top2vec is based on the assumption that the joint word and document embedding is a semantic embedding that learns the semantic association between words and documents \cite{DBLP:journals/corr/abs-2008-09470}. Moreover, the semantic space can be seen as a continuous representation of topics where a dense cluster of points can be interpreted as documents pertaining to a similar topic. Top2vec creates jointly embedded topic, document, and word vectors with semantic similarity being the distance between them. To account for document vector sparsity, dimensionality reduction with UMAP (Uniform Manifold Approximation and Projection) is utilized \newcite{mcinnes2020umap}. HDBSCAN which is a density-based clustering algorithm \cite{McInnes2017} is then used to find dense clusters of documents. The author reports that Top2vec topics are more informative and offer a better representation of the corpus, compared to those offered by LDA and PLSA.
The recent topic classification research has been steering towards neural networks and pretrained language models \cite{10.1145/3372938.3372951,DBLP:conf/mie/DanilovIKOSP21}. BERTopic is another recent technique \newcite{grootendorst2020bertopic}, very similar with Top2vec. It makes it possible to use BERT \cite{devlin-etal-2019-bert} variants for generating document embeddings, and same as Top2vec utilizes UMAP and HDBSCAN for dimensionality reduction and clustering. It further uses a class-based TF-IDF (applying TF-IDF on clusters of documents) to find the important topic words and then the topics. Unfortunately, there is still no comparison between Top2vec and BERTopic.  

\section{Topic Modeling Procedure}
\label{sec:topicmod}      

To perform topic modeling and segmentation on the OAG collection, we utilized LDA and Top2vec.\footnote{BERTopic computation requirements could not be met.} For constructing a highly-representative taxonomy of topics, we performed several parameter optimizations on both methods. Since we also wanted to have a collection of realistic data samples, we also performed two filtering steps that removed the outliers (very long or very short record fields). The details of these data processing steps are given in the following sections.    

\subsection{Data Preprocessing}
\label{ssec:datapreproc}

We started from OAG version 2.1 and applied a first filtering step to drop out all article records without \emph{title}, \emph{abstract} and \emph{keywords} (three fields available in every sample of OAGT). Authors names were intentionally removed to avoid any privacy concerns. Using the \emph{language} attriburte, we also removed records not in English. All other article attributes like \emph{publisher}, \emph{year}, \emph{venue}, \emph{volume}, \emph{issue}, \emph{isbn}, \emph{issn}, \emph{doi}, \emph{url}, etc. were retained in every sample they appeared. These steps reduced the collection to about 15 million samples, from about 31 million that were initially retrieved.
After completing the above initial preparatory steps, we explored some of the samples. There were outliers in certain fields (e.g., very long titles or abstracts). We used CoreNLP tokenizer \cite{manning-etal-2014-stanford} and removed records with a title not within 2 - 40 tokens and abstracts, not in the range of 70 - 450 tokens. Since author keywords played an important role in the process, we also removed records with fewer than 3 and more than 35 of them. We also tried to keep the collection clean of non-standard symbols (not in the ASCII range of UTF-8). Records with more than five such symbols in abstract and title (joined together) were removed. After this second set of filtering steps, the collection was reduced to 6\,942\,930 records which is also the final size of OAGT.

\subsection{Data Preparation}
\label{ssec:dataprep}

To facilitate the memory requirements of the topic modeling methods, we used a ``shortcut''. Instead of analyzing combinations of title, abstract, and author keywords, we used comma-separated keyword hashes made up of the authors' keywords enriched with other keyterms extracted from the joined abstract and title string. This way, we are assured that the overall data sample topicality captures the influence of title, abstract, and author keywords together.
To extract these extra keyterms, we utilized an implementation\footnote{\url{https://pypi.org/project/rake-nltk}} of RAKE, a popular method for keyword extraction that is fast, domain-independent, and language-independent \cite{doi:https://doi.org/10.1002/9780470689646.ch1}. The keyterms extracted with RAKE were limited to a length of 1 to 3 (unigrams, bigrams, and trigrams only). An example of such article keyword hash is shown in Table~\ref{tab:kwhash}.
\begin{table}[ht]
\begin{center}
\begin{tabular}{|p{70mm}|}
\hline
cell cycle , dna binding proteins , signal transduction , dna damage , diploidy , dna replication , genetic loci , dna repair , heterozygote , mutation , replication stress presented , rad6 dependent pathways , particular cell types \\
\hline
\end{tabular}
\caption{\label{tab:kwhash}Example of a keyword hash}
\end{center}
\end{table} 
Considering that there was an average of 8 author keywords per sample, we tried to reach a roughly uniform keyword hash size. To this end, we varied the number of extra keyterms extracted with RAKE based on the number of author keywords for each sample (more extra keyterms for fewer author keywords and vice-versa). In the end, the average keyword hash size of the entire collection was roughly 14 tokens. 

\subsection{Topic Modeling Optimizations} %
\label{ssec:lda}

The quality of topic modeling depends on several parameters which require optimization. One way to optimize those model parameters is by trying different values for each of them to maximize the topic coherence score, a measure of the degree of semantic similarity between high scoring words in the respective topics \cite{10.1145/2684822.2685324}. We used the topic coherence to optimize different LDA model parameters. 
Starting with the dictionary, we removed words that are very common (more frequent than the value of \texttt{no\us above} parameter) or very rare (less frequent than the \texttt{no\us below} parameter). In the case of \texttt{no\us below}, we tried all values from 10 to 150 and found 105 (words appearing less than 105 times discarded) to be the optimal one. For \texttt{no\us above}, we tried values $0.2, 0.3, ..., 0.9$ and found 0.4 (words with frequency higher than 40\,\% discarded) to be the optimum.
Two more model parameters that we optimized are \texttt{iterations} and \texttt{passes}. The former dictates the number of iterations of the model on each document of the collection, whereas the latter dictates the number of iterations on the entire document collection (this is known as \texttt{epochs} in the context of neural networks). Normally, higher values of these parameters could lead to better models, but would also require more memory. The best we could do within our limited computing resources was trying values $1, 2, ..., 5$ for each of them. The optima we found were 5 and 3 for \texttt{iterations} and \texttt{passes} respectively. 
The final optimization we performed was about the number of topics (\texttt{num\us topics} parameter). We tried all values between 5 and 150 and got 27 as the optimum, once again with respect to maximizing the value of the topic coherence score.    
Compared to LDA, Top2vec offers the advantage of performing certain optimizations automatically. Since it handles frequent workds (and especially stop words) by itself, there is no \texttt{no\us above} to set or optimize. Top2vec offers the possibility to chose between \texttt{fast-learn} (runs fast but produces lower-quality vectors), \texttt{learn} (balance between run speed and vector quality), and \texttt{deep-learn} (best quality of vectors, but runs slowly). We worked with \texttt{learn} option, since \texttt{deep-learn} was taking too long and could even fail because of out-of-memory errors. For the other parameters like \texttt{embedding\us model} (using doc2vec by default), \texttt{chunk\us length} (length of document chunks), \texttt{embedding\us batch\us size}, etc. we used the default values.     

\subsection{Taxonomy of Topics}
\label{ssec:taxonomy}

\begin{table*}[ht]
\begin{center}
\begin{tabular}{|p{152mm}|}
\hline
"title": "Healthcare spending and health outcomes: evidence from selected East African countries.", "abstract": "The results of this study have important policy and management implications for the eight East African countries. From a policy perspective, it is necessary to understand if a greater allocation of resources to the healthcare sector is worthwhile and to determine whether to encourage private healthcare investment. From the management perspective, investing in more private institutions, such as hospitals and clinics, is essential for health outcomes in the average country. The results of this study can be used by the World Health Organization as well as other non-governmental organizations that provide financial assistance to East African countries.", "keywords": ["Healthcare expenditures", "health outcome", "infant deaths", "life expectancy", "neonatal deaths", "under-five deaths"], "lang": "en", "volume": "17", "issue": "1", "issn": "1729-0503", "topic": 6, "topic\_terms": ["respondents", "mental", "questionnaires", "computed tomography", "prevention", "interviews", "care", "occupational", "brain mapping", "ergonomics", "eye movement", "suicide", "health", "human factors", "occupational safety", "injury prevention", "suicide prevention", "social network", "dsp", "alcohol"] \\
\hline
\end{tabular}
\caption{\label{tab:oagtsamp}A data sample example from OAGT dataset}
\end{center}
\end{table*}
\begin{table}[ht]
\begin{center}
\begin{tabularx}{\columnwidth}{|l|X|}
\hline
\bf Attribute & \bf Value \\
\hline
Title length in tokens & 9.75\,(2.67) \\
\hline
Abstract length in tokens & 149.84\,(59.14) \\
\hline
Author keywords & 8.09\,(5.21) \\
\hline
Keywords-Topics Jindex & 0.4\,\%\,(1.3\,\%) \\
\hline
Keywords-Topics overlap & 0.72\,\%\,(2.3\,\%) \\
\hline
\end{tabularx}
\caption{\label{tab:oagtstats}Statistics of OAGT dataset}
 \end{center}
\end{table}
We combined the top 10 topic words from LDA and Top2vec respectively (first word from Top2vec, second from LDA, third from Top2vec, etc.) and formed the taxonomy of 27 topics, with 20 terms for each topic. It is presented in Table~\ref{tab:oagtterms}. The combination of topic words preserved their order (most significant first) as returned from each method. We carefully replaced duplicated terms (with the next high-ranking term) by performing stemming. The duplicate terms were mostly plural forms such as ``control system'' and ``control systems'', ``pixel'' and ``pixels'', etc.  
One thing we observed is that Top2vec tends to yield single-word phrases instead of n-grams. Contrary, LDA produces more bigrams and trigrams. Overall, the topic terms are mostly unigrams.    
As for the topic size (number of records pertaining to each topic), we can see from Table~\ref{tab:oagtterms} that the biggest is \texttt{topic0} with 401\,338 documents and the smallest is \texttt{topic26} with 147\,896 documents (topics are actually ranked by their size). 
The main benefit of the segmentation process is the possibility to use OAGT in two ways: either as a huge heterogeneous dataset of multiple topics or as a collection of 27 datasets, each being a homogeneous corpus with samples from a specific research domain.  

\section{OAGT Dataset}	
\label{sec:oagmt}

Table~\ref{tab:oagtsamp} shows a data sample example from OAGT. We computed some basic statistics from the dataset which are presented in Table~\ref{tab:oagtstats}. As we can see, the average length as the number of tokens for the title and abstract is 9.75 and 149.84 respectively. The number of authors' keywords is roughly 8.  
We further tried to check the similarity between author keywords and topic terms. One way to do this is by computing the Jaccard similarity (Keywords-Topics Jindex in Table~\ref{tab:oagtstats}) between the two sets of tokens using the following equation:
\begin{equation*}
\label{eq:jacardindex}
J(A, T) = \frac{|T \cap A|}{|T \cup A|} = \frac{|T \cap A|}{|T| + |A| - |T \cap A|}
\end{equation*}
where $T$ is the set of unique tokens in the topic terms and $A$ is the set of unique tokens in the authors' keywords. The average value we computed for this metric in the whole OAGT is 0.4\,\%. Another metric that shows the similarity between author keywords and the topic terms is the overlap $o(a, t) = \frac{|\{a\} \cap \{t\}|}{|\{t\}|}$ which computes the fraction of unique topic tokens $t$ that overlap with a token from authors' keywords $a$. The average value of this overlap in the whole OAGT is 0.72\,\%.
These statistics indicate that the similarity between the authors' keywords and the topic terms is too low. We carefully checked some records and noticed that indeed, the authors' keywords are specific and tightly related to the title and abstract. Contrary, the topic terms are way more generic and provide only a high-level overview of what that paper is about.

\section{Discussion} %
\label{sec:discussion}

Many large collections of research article data that are available today can be used for experimenting on tasks such as named entity recognition, text summarization, topic analysis, keyword generation, etc. One problem with these available datasets is the fact that they are heterogeneous and topically unstructured, in the sense that the articles they contain belong to various research disciplines. The few homogeneous datasets with records from a single discipline are relatively small. 
In this work, we created OAGT, a dataset of 6\,942\,930 research article data records that is both large and multitopic. We combined LDA and Top2vec, two popular topic modeling methods to construct a taxonomy of 27 topics. Each topic of the taxonomy is represented by its 20 terms which are the most significant ones. 
The topic annotation of each record makes it possible to use this data collection in two modalities: as a huge heterogeneous set of multiple topics (same as most existing collections, ignoring the topic annotations we added) or as 27 large sets (the smallest having 147\,896 documents) of samples from different topics, each set being homogenous, with documents of a single topic.
One serious difficulty we faced was the high computation requirements when working with data collections of this size. As a result, we couldn't utilize other advanced topic recognition techniques such as BERTopic for reaching an even better taxonomy of topics. 
Another limitation of our work is the lack of explicit topic identifiers attached to each record. One would expect to have each document annotated by a hierarchical scheme of 1 -- 3 levels of terms, same as those used by editors or publication libraries (e.g., a paper about immunology research could be labeled \texttt{medicine.immunology}).
In the future, we plan to overcome this limitation by trying different approaches. One possibility is to integrate knowledge from online classification systems of research works such as that of ACM.\footnote{\url{https://dl.acm.org/ccs}} Another possibility is to hire experts who could manually annotate hundreds of records, and then utilize semi-supervised learning techniques \cite{10.1145/2983323.2983752,YAKIMOVICH2021100383} for scaling up to the entire collection.   
\section{Bibliographical References}\label{reference}

\bibliographystyle{lrec2022-bib}
\bibliography{lrec2022-example}

\section*{Appendix: Table with topic terms}

The list of topics and their terms, together with the number of documents pertaining to each topic are presented in Table~\ref{tab:oagtterms}.

\begin{table*}[ht]
\begin{center}
\scriptsize
\begin{tabular}{|c|c|p{135mm}|}
	\hline
	\bf Topic & \bf Docs & \bf Terms \\
	\hline
	0 & 401338 & 'ideas', 'reform', 'universities', 'mental health', 'enlightenment', 'research methodology', 'innovation', 'connotation', 'teachers', 'innovate', 'innovating', 'predicament', 'evaluation', 'developing countries', 'population', 'behavior', 'health', 'stroke', 'quality', 'culture' \\
	\hline
	1 & 365225 & 'user', 'web', 'developers', 'web services', 'desktop', 'groupware', 'virtual reality', 'developer','information system', 'browser', 'usability', 'users', 'pages', 'internet', 'information technology','web service', 'cognition', 'memory', 'case study', 'world wide web' \\
	\hline
	2 & 363616 & 'phonons', 'electrons', 'superlattices', 'photoemission', 'debye', 'dielectric', 'electrodynamics', 'human computer interaction', 'resonances', 'kerr', 'semiconducting', 'optimization', 'children', 'frequency', 'identification', 'voltage', 'learning', 'silicon', 'review', 'switches' \\
	\hline
	3 & 324116 & 'pixels', 'histograms', 'image', 'algorithm design analysis', 'segmentation', 'noisy', 'statistical analysis', 'grayscale', 'thresholding', 'denoising', 'computer vision', 'multiresolution', 'gaussians', 'data mining', 'feature extraction', 'image processing', 'image segmentation', 'computational complexity', 'learning artificial intelligence', 'wireless sensor network' \\
	\hline	
	4 & 322747 & 'aqueous', 'oxides', 'inorganic', 'transmission electron microscopy', 'photocatalysis', 'x ray diffraction', 'xps', 'titania', 'xrd', 'tio', 'sio', 'nanocomposite', 'iron', 'copper', 'nitrogen', 'scanning electron microscopy', 'kinetics', 'zinc', 'adsorption', 'polymer' \\
	\hline	
	5 & 305815 & 'torque', 'fuzzy logic', 'simulink', 'mobile robots', 'servomotors', 'fuzzy control', 'excavator', 'decision making', 'process control', 'brushless', 'braking', 'torques', 'maneuverability', 'vehicles', 'motors', 'real time', 'control system', 'navigation', 'safety', 'robots' \\
	\hline	
	6 & 283064 & 'respondents', 'mental', 'questionnaires', 'computed tomography', 'prevention', 'interviews', 'care', 'occupational', 'brain mapping', 'ergonomics', 'eye movement', 'suicide', 'health', 'human factors', 'occupational safety', 'injury prevention', 'suicide prevention', 'social network', 'dsp', 'alcohol' \\
	\hline
	7 & 280036 & 'leaf', 'cultivars', 'herbaceous', 'foliar', 'woody', 'plants', 'solar radiation', 'breeding', 'flowering', 'crop', 'grasses', 'ornamental', 'varieties', 'pollinated', 'fruiting', 'agronomic', 'vegetative', 'seed', 'harvest', 'eggplant' \\
	\hline	
	8 & 276952 & 'transduction', 'cell proliferation', 'signaling', 'nitric oxide', 'pathways', 'overexpression', 'knockdown', 'immune response', 'cytoskeletal', 'tumorigenesis', 'phosphorylation', 'proliferation', 'cell', 'apoptosis', 'cancer', 'cell cycle', 'immunohistochemistry', 'aging', 'cell death', 'oncology' \\
	\hline	
	9 & 269196 & 'solvability', 'numerical simulation', 'nonsingular', 'differential equation', 'functionals', 'adjoint', 'numerical method', 'equations', 'theorems', 'cauchy', 'mathematical models', 'algebraic', 'asymptotic', 'infinity', 'three dimensional', 'mathematical model', 'finite element analysis', 'computer simulation', 'finite element', 'monte carlo method' \\
	\hline	
	10 & 268964 & 'semantics', 'embedded system', 'formalisms', 'software systems', 'languages', 'formal specification', 'formalization', 'semantical', 'programming language', 'syntactical', 'operating system', 'logical', 'predicate', 'software architecture', 'abstractions', 'specification', 'software development', 'distributed system', 'cloud computing', 'object oriented programming' \\
	\hline	
	11 & 264696 & 'multihop', 'mobile communication', 'retransmissions', 'resource allocation', 'unicast', 'hop', 'wireless communication', 'internet', 'wireless', 'packet', 'routing', 'multicast', 'forwarding', 'multiaccess', 'quality service', 'protocols', 'scheduling', 'throughput', 'ad hoc networks', 'bandwidth' \\
	\hline	
	12 & 251168 & 'overfitting', 'reinforcement learning', 'clustering', 'mathematical programming', 'dimensionality', 'outperform', 'global optimization', 'bayesian', 'rule based', 'supervised', 'measurement system', 'classifiers', 'algorithms', 'unsupervised', 'datasets', 'knowledge representation', 'pattern matching', 'training data', 'bayesian network', 'cross correlation' \\
	\hline	
	13 & 246896 & 'alloy', 'cooling', 'heavy metals', 'metallurgy', 'energy dissipation', 'aluminum', 'preheating', 'thermal', 'alloying', 'steel', 'stainless', 'temperature', 'microstructure', 'furnace', 'heating', 'carburization', 'molten', 'abrasive', 'corrosion', 'welding' \\
	\hline	
	14 & 245648 & 'complication', 'body weight', 'endovascular', 'control group', 'hemodynamic', 'blood pressure', 'artery', 'young adult', 'femoral', 'risk factors', 'surgery', 'quality life', 'fractures', 'orthopedic', 'anatomy', 'trauma', 'diagnosis', 'obesity', 'treatment', 'prevalence' \\
	\hline	
	15 & 241685 & 'hydrology', 'energy efficiency', 'basins', 'rivers', 'water resources', 'geomorphology', 'environmental management', 'coastal', 'continental', 'energy conservation', 'geology', 'climatology', 'energy consumption', 'renewable energy', 'sedimentology', 'ocean', 'climate', 'sea', 'lakes', 'sediments' \\
	\hline	
	16 & 231388 & 'market', 'population dynamics', 'profits', 'economic development', 'prices', 'developing countries', 'investment', 'economies', 'finance', 'macroeconomics', 'microeconomics', 'capital', 'revenues', 'liquidity', 'revenue', 'taxation', 'management', 'sustainable development', 'population', 'statistics' \\
	\hline	
	17 & 224431 & 'genome', 'genetic diversity', 'dna', 'pseudogenes', 'genetic markers', 'transposable', 'primers', 'genetics', 'genotype', 'genetic variation', 'chromosomes', 'gene', 'nucleotide', 'molecular genetics', 'alleles', 'structural change', 'putative', 'globins', 'population genetics', 'nucleotides' \\
	\hline	
	18 & 223377 & 'reinforced', 'bending', 'prestressed', 'displacement', 'abaqus', 'elastoplastic', 'deformation', 'thermal analysis', 'fem', 'ductility', 'flexural', 'crack', 'prestress', 'deflection', 'temperature field', 'girder', 'girders', 'prestressing', 'shear', 'loading' \\
	\hline	
	19 & 217487 & 'attentional', 'cognition', 'verbal', 'cognitive development', 'perception', 'psychology', 'cognitive', 'neuropsychology', 'fault diagnosis', 'visuospatial', 'emotional', 'nonverbal', 'emotion', 'metacognition', 'perceptual', 'mentalizing', 'fluency', 'participants', 'stimuli', 'dyslexia' \\
	\hline	
	20 & 212891 & 'stereoisomerism', 'enzyme', 'intermediates', 'enzyme activity', 'amino', 'chemical composition', 'biocatalysis', 'sulfhydryl', 'enzymic', 'conformation', 'nuclear medicine', 'metalloproteins', 'peroxidases', 'monomeric', 'hydrolysis', 'coenzymes', 'acids', 'kinetics', 'enzymatic', 'oxidoreductases' \\
	\hline	
	21 & 210400 & 'diet', 'fasting', 'obesity', 'clinical trials', 'metabolic', 'insulin', 'drug delivery', 'supplementation', 'dyslipidemia', 'candidate gene', 'vitamin', 'cardiometabolic', 'hyperglycaemia', 'diabetes', 'dietary', 'glycemic', 'multifactorial', 'triglycerides', 'albuminuria', 'appetite' \\
	\hline	
	22 & 208871 & 'processors', 'multiprocessing', 'signal detection', 'overheads', 'transmission line', 'microprocessors', 'multicore', 'channel capacity', 'multiprocessors', 'parallelism', 'benchmarks', 'communication system', 'execution', 'signal processing', 'compilers', 'signal analysis', 'superscalar', 'architectures', 'cpu', 'pipelining' \\
	\hline	
	23 & 200184 & 'resection', 'metastases', 'staging', 'colorectal cancer', 'chemoradiation', 'recurrence', 'oncologic', 'metastatic', 'metachronous', 'locoregional', 'neoplasm', 'lymphadenectomy', 'public policy', 'preoperative', 'sarcomatoid', 'radiotherapy', 'microfluidics', 'malignant', 'oncological', 'tumors' \\
	\hline	
	24 & 178073 & 'gabaergic', 'neurochemical', 'gaba', 'tissue engineering', 'glutamatergic', 'excitatory', 'neurons', 'neuronal', 'striatum', 'interneurons', 'myocardial infarction', 'thalamus', 'neocortex', 'presynaptic', 'spinal cord', 'excitability', 'interneuron', 'cholinergic', 'bicuculline', 'striatal' \\
	\hline	
	25 & 176770 & 'pathogen', 'escherichia coli', 'virulence', 'serotypes', 'biological activity', 'serology', 'virulent', 'dna sequence', 'serotype', 'coinfection', 'outbreaks', 'bacterial', 'isolates', 'antigenic', 'infect', 'cell culture', 'virus', 'amino acids', 'vaccines', 'viral' \\
	\hline	
	26 & 147896 & 'countries', 'demographic', 'socioeconomic', 'state estimation', 'married', 'oceania', 'america', 'migrants', 'critique', 'heterosexuals', 'unmarried', 'attitudes', 'statutes', 'parenthood', 'midwives', 'sexuality', 'marital', 'motherhood', 'religion', 'youth' \\
	\hline
\end{tabular}
\caption{\label{tab:oagtterms}Topic terms and number of samples for each topic}
\end{center}
\end{table*}

\end{document}